# Human Genome Book:Words,Sentences and Paragraphs

Wang Liang*
Huazhong University of Science and Technology, 430070, P.R. China
*To whom correspondence should be addressed. E-mail:wangliang.f@gmail.com

[Abstract]Since the completion of the human genome sequencing project in 2001, significant progress has been made in areas such as gene regulation editing and protein structure prediction. However, given the vast amount of genomic data, the segments that can be fully annotated and understood remain relatively limited. If we consider the genome as a book, constructing its equivalents of words, sentences, and paragraphs has been a long-standing and popular research direction. Recently, studies on transfer learning in large language models have provided a novel approach to this challenge.Multilingual transfer ability, which assesses how well models fine-tuned on a source language can be applied to other languages, has been extensively studied in multilingual pre-trained models. Similarly, the transfer of natural language capabilities to "DNA language" has also been validated. Building upon these findings, we first trained a foundational model capable of transferring linguistic capabilities from English to DNA sequences. Using this model, we constructed a vocabulary of DNA words and mapped DNA words to their English equivalents.Subsequently, we fine-tuned this model using English datasets for paragraphing and sentence segmentation to develop models capable of segmenting DNA sequences into sentences and paragraphs. Leveraging these models, we processed the GRCh38.p14 human genome by segmenting, tokenizing, and organizing it into a "book" comprised of genomic "words," "sentences," and "paragraphs." Additionally, based on the DNA-to-English vocabulary mapping, we created an "English version" of the genomic book. This study offers a novel perspective for understanding the genome and provides exciting possibilities for developing innovative tools for DNA search, generation, and analysis.

# 1 Introduction

Since the completion of the first human genome map by the Human Genome Project in 2001, significant advancements have been made in fields such as gene regulation editing and protein structure prediction. However, given the vast amount of genomic data, the segments that can be fully annotated and deeply understood remain relatively limited. To achieve a more comprehensive understanding of genomic information, many research efforts have begun exploring the possibility of treating the genome as a book, aiming to construct its "words," "sentences," and "paragraphs," similar to the parsing of human language (1-3).

Most of these studies focus on analyzing the statistical similarities between DNA sequences and natural language, including functional comparisons between potential basic units of DNA (e.g., words) and those in natural language (4-8). However, due to the substantial differences between natural language and biological sequences, most of these studies remain experimental and analytical in nature, making it challenging to develop effective methods or theories. For example, transferring the logical reasoning and summarization capabilities of natural language models to DNA sequences has proven difficult. Consequently, these studies still fall short of constructing a genomic "book" with the clarity of structure and explicit meaning found in natural language books.

In recent years, however, advances in large language models have provided entirely new tools for comprehensively interpreting genomic data.

The rise of large language models has revolutionized artificial intelligence, significantly impacting fields like bioinformatics. For nucleic acid analysis, models such as DNABert2, HyenaDNA, and EVO have been developed to address challenges in DNA sequence classification and structural prediction (9–13). By harnessing the capabilities of these advanced models, researchers are gaining fresh perspectives on genetic information.

In parallel, advancements in protein studies have introduced tools like ProTrans, ProteinBERT, and ESM2. These models are tailored for tasks such as protein structure prediction and functional annotation (14-24). Together, these breakthroughs demonstrate the adaptability of large language models in tackling intricate biological problems, fostering a deeper connection between computational linguistics and molecular biology.

An intriguing direction within these studies is the exploration of multilingual transfer capabilities in large language models. Research on multilingual transfer has demonstrated how large pre-trained models can effectively apply knowledge learned in one source language to other languages. These studies have validated the effectiveness of cross-linguistic knowledge transfer, including between different natural languages, as well as between programming languages and natural language (25-29). Moreover, some recent experimental research has shown that the transfer phenomenon from natural language capabilities to DNA sequences also exists, making it possible to directly apply natural language processing techniques to DNA sequence analysis (30).

In this paper, we leverage the transfer of natural language capabilities to DNA language to construct a structured human genomic "book." Specifically, we pre-trained a GPT-2 model, **gpt2-gene-eng**, on English, DNA, and protein sequences using a unified BPE tokenizer. We then fine-tuned this model using the English semantic similarity dataset from PAWSX, resulting in a model, **gpt2-gene-eng-ft**, capable of transferring natural language abilities to DNA sequences. Based on this fine-tuned model, we further trained three new models using English datasets for sentence splitting, paragraph segmentation, and summarization tasks, respectively. These three models were subsequently applied to process human genome data, producing a genomic "book" that includes DNA words, sentences, and paragraphs.

Additionally, by utilizing the pre-trained **gpt2-gene-eng** model, we established a mapping between the DNA vocabulary and the English vocabulary, enabling the creation of an English-translated version of the human genomic book.

## 2 Materials and methods

### 2.1 Overview of the Genome Book Construction Process

Leveraging the language transfer capabilities of large language models, the construction of a genome book is divided into the following steps:

**1 Construction of the Pre-trained Model gpt2-gene-eng**

A pre-trained model was built using English data, DNA sequences, and protein sequences, with a unified BPE tokenizer. The base model architecture used in this study is GPT-2.

**2 Enabling Language Transfer to Create gpt2-gene-eng-ft**

The model was fine-tuned on English similarity judgment datasets to enable the transfer of English language capabilities to DNA. Sequence similarity judgment tasks are one of the few tasks that can validate the transfer of natural language abilities to DNA sequences. This step produced the gpt2-gene-eng-ft model.

**3 Further Fine-tuning for Sentence Splitting, Paragraph Segmentation, and Summarization**

Based on gpt2-gene-eng-ft, the model was further fine-tuned using English datasets for sentence splitting, paragraph segmentation, and summarization, resulting in three new models with specialized capabilities. Since punctuation such as the period (".") was already included in the pre-training data, sentence splitting can be accomplished by using gpt2-gene-eng-ft to predict the "." token. The paragraph segmentation model, named *gene_eng_gpt2_para_seg*, and the summarization model, named *gene_eng_gpt2_summary*, were also developed during this step.

**4 Processing Genome Data to Generate the Genome Book**

Using the aforementioned models, genome data was processed to create a genome book containing "words," "sentences," and "paragraphs," as illustrated in the following diagram:

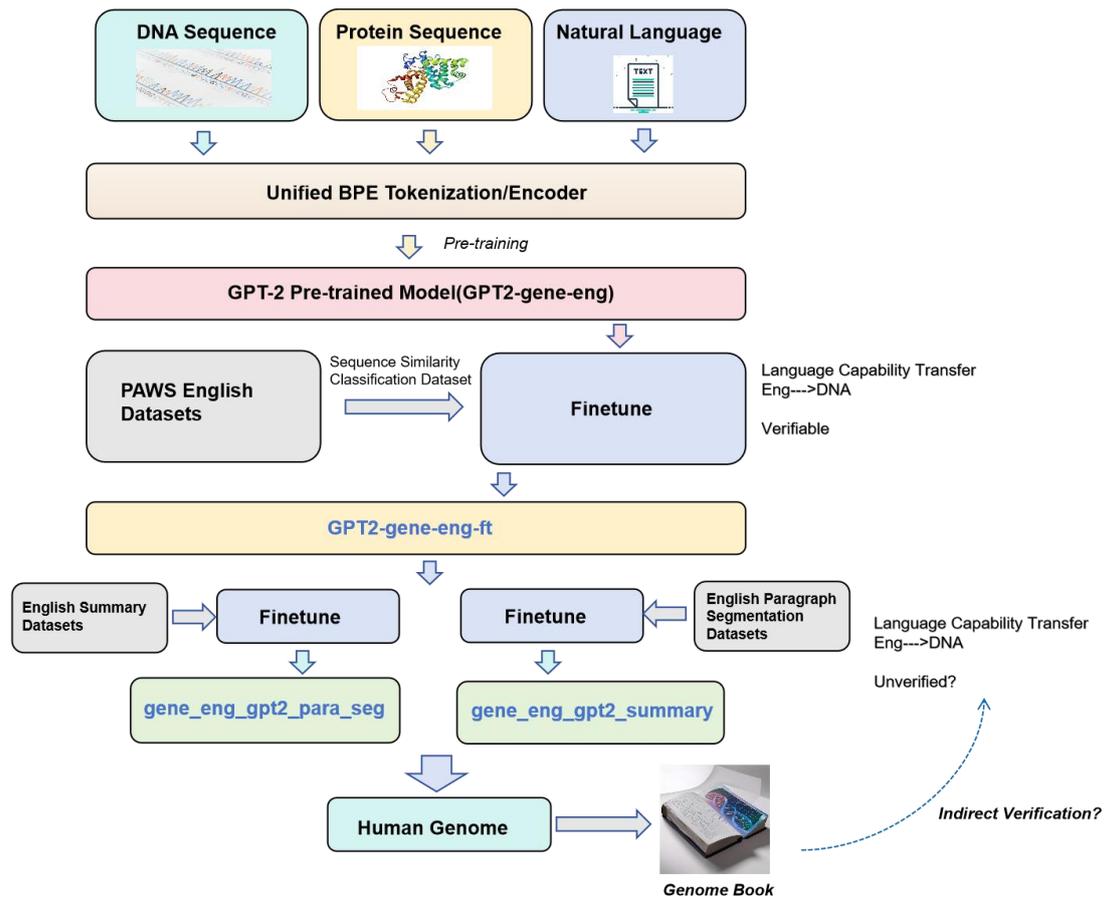

**Fig.1. Construction Process of the Human Genome Book.** First, a model, GPT2-gene-eng-ft, capable of transferring natural language abilities to DNA language was developed. This step involves verifiable language transfer. Subsequently, this model was further fine-tuned to acquire DNA-specific abilities for segmentation, sentence splitting, and summarization. It is important to note that these abilities were trained using relevant English datasets, and their transfer to DNA sequences cannot be directly verified. We hypothesize that the abilities for segmentation, sentence splitting, and summarization can be transferred to DNA sequences. Based on this assumption, the human genome data was processed to construct a genome book containing "words," "sentences," and "paragraphs." If novel and effective DNA research tools or new biological phenomena can be discovered using this genome book, it would provide indirect validation of the existence of these transferred abilities.

## 2.2 Pretrained model

The training data for is roughly outlined in the table below:

Table.1 Training data

| period | datasets | data size | Data type |
| --- | --- | --- | --- |
| BPE training | multiple model organism genomes | 2G | DNA |
| BPE training | UniProt [multispecies] | 2G | Protein sequence |
| BPE training | OpenWebText | 2G | English |

| pre training | multiple model organism genomes | 10G | DNA |
| pre training | UniProt+Swiss-Prot/TrEMBL | 10G | Protein sequence |
| pre training | Wiki english+OpenWebText | 10G | English |

The training data for the paper is structured as follows:

**DNA Sequence Data**:

We followed the pre-training data approach used by DNABERT, extracting fragments ranging from 300 to 1000 base pairs (bp) from multiple model organisms. The total volume of DNA sequence data is approximately 10 GB, with 2 GB of this data randomly selected for training the tokenizer.

**Protein Sequence Data**:

From the UniProt database, we extracted 10 GB of protein sequence data, including all entries from Swiss-Prot and randomly selected entries from TrEMBL.

**Natural Language Data**:

We primarily used the OpenWebText dataset, combined with Wikipedia data, as the natural language training dataset, focusing on English text. From these datasets, we extracted 10 GB of data for pre-training purposes.

One important condition for achieving multilingual transfer ability is the use of a unified tokenizer, typically based on the Byte Pair Encoding (BPE) method. In this study, we trained a BPE tokenizer from scratch using DNA, protein, and English data, resulting in a vocabulary of approximately 100,000 tokens.

For the model, we employed the GPT2-small architecture and trained a GPT2 model from scratch using DNA, protein, and English data, referred to as **gpt2-gene-eng**. The specific network structure is detailed as follows:

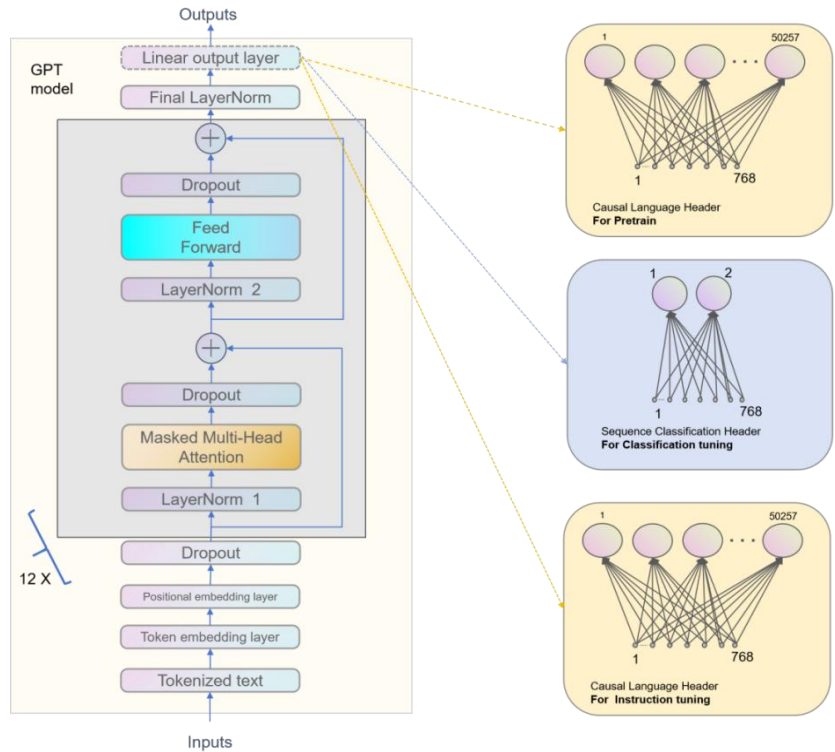

Fig.2. Structure of **gpt2-gene-eng**. The model architecture is identical to that of GPT-2 small. Pre-training and instruction tuning use the same causal language header, the layer mapped 768 hidden units to 100000 units (the number of tokens in the vocabulary). Classification uses sequence classification header, The layer maps from 768 hidden units to only 2 units, where the 2 units represent the two classes id.

According to the typical design of GPT-2, it accepts sequences with a maximum length of 1024 as input. We used the same architecture as the GPT-2 Small model, which consists of 12 Transformer layers, each with 768 hidden units and 12 attention heads. This small model has approximately 117 million parameters. We trained the GPT-2 model using mixed-precision floating-point arithmetic on a machine equipped with a single Nvidia 4090 GPU. We employed a dynamic learning rate schedule, and the model was trained for a total of 3 to 5 epochs.

**2.3 Finetune model with multilingual transfer ability**

Following the approach described in reference (30), we fine-tuned the pre-trained model using the English semantic similarity task from the PAWSX dataset to enable language transfer capabilities. The fine-tuning process utilized a classification head, and the fine-tuned model is denoted as **gpt2-gene-eng-ft**.

Additionally, we employed the DNA similarity sequence judgment dataset from the same study to validate the language transfer capability. The results of testing the transfer of English language capabilities to DNA in the **gpt2-gene-eng-ft** model are as follows:

Table.2 Accuracy of different datasets

| model | pretrain | finetune | test Dna150s | test Dna150 | test Dna50 |
|---|---|---|---|---|---|
| gpt2-gene-eng-ft | DNA+eng+protein | en | 0.92 | 0.79 | 0.86 |

We began by performing classification fine-tuning on the English sequence similarity dataset from PAWS-X, followed by testing on DNA datasets.

DNA150s, DNA150, and DNA50 are DNA sequence similarity judgment datasets constructed with varying lengths and strategies. The model fine-tuned on the English semantic similarity dataset was tested on these DNA datasets, achieving an accuracy of over 79% across all cases. This indicates that English language capabilities can indeed be transferred to DNA language.

To visualize this more clearly, we employed PCA for dimensionality reduction and plotted the DNA word vectors and English word vectors in a two-dimensional graph. The specific visualization steps are as follows:

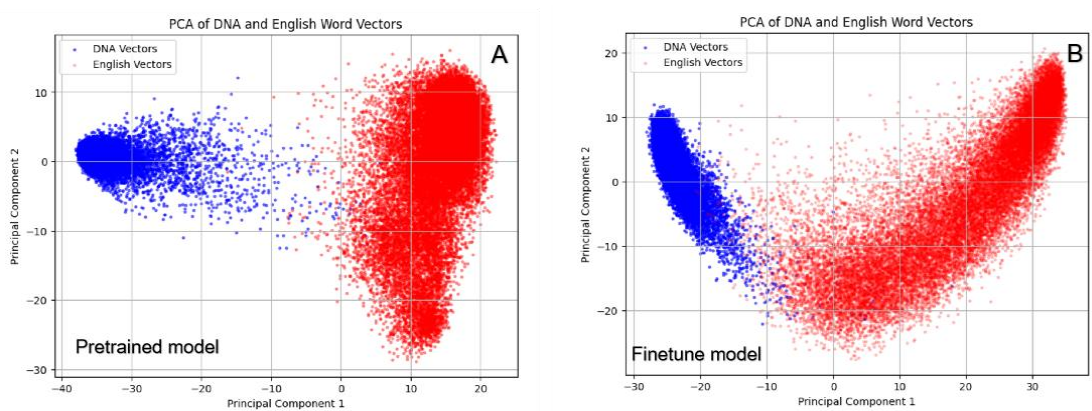

Fig.3.On the left, Figure A shows the word vectors generated by the pre-trained model. Blue points represent DNA word vectors, and red points represent English word vectors. On the right, Figure B shows the word vectors generated by the fine-tuned model. Again, blue points represent DNA word vectors, and red points represent English word vectors.In Figure A, it is evident that the DNA word vectors and English word vectors are generally far apart, with a clear boundary between them. In contrast, in Figure B, the DNA word vectors and English word vectors are noticeably closer, with some overlapping regions. This further illustrates the reason for language capability transfer in the fine-tuned model.

## 2.4 Text Segmentation Model

For the task of text segmentation, two primary strategies are generally adopted:

**1. Direct Prediction of Paragraph Markers**: Train the model to predict whether the next token is a paragraph marker after each token.

**2. Binary Classification Problem**: Transform the task into a binary classification problem, where the model determines if a given position marks a paragraph boundary. This involves adding an extra classification layer after each token, outputting 0 or 1 to indicate the absence or presence of a paragraph boundary.

We opted for the relatively simpler Strategy 1.

**Training Data**

To develop a model suitable for segmentation tasks, we primarily utilized Wikipedia as our data source. Wikipedia data stands out for its rich content, clear paragraph structure, and broad thematic coverage, making it ideal for studying paragraph prediction. Initially, we preprocessed the raw Wikipedia text by replacing newline characters between paragraphs with a specific marker <p_end> to clearly denote the end of a paragraph. After this preprocessing, the boundaries of paragraphs in each article within the dataset were uniformly presented, aiding the model in learning the patterns of paragraph endings. Additionally, to prevent data bias, we imposed a limit on text length, ensuring that each sample contained multiple segments while preserving contextual information.

**Model Training**

We used the GPT2LMHeadModel with a causal language modeling head as the base model for paragraph prediction. The training process was based on a language modeling task, employing continuous pre-training to enable the model to learn the patterns of paragraph boundaries. Post-preprocessing, texts marked with <p_end> were used as training inputs, with the model's objective being to predict the next token, including both regular vocabulary and the paragraph end marker <p_end>. Through this approach, the task of predicting paragraph boundaries was integrated into the language modeling task without the need to design additional task heads.

During training, we employed cross-entropy as the loss function and introduced a dynamic learning rate adjustment strategy to optimize convergence speed. We chose the AdamW optimizer with an initial learning rate set at $5\times10^{-5}5\times10^{-5}$. The model learned the patterns of paragraph boundaries on the training set while evaluating the accuracy and fluency of generated text on the validation set. After several rounds of training, the model was able to capture contextual information and accurately generate paragraph end markers.

## 2.5 Sentence Boundary Detection Model

In the sentence segmentation task, we adopted a strategy identical to that used for paragraph segmentation. Since the pre-trained model's data already includes punctuation marks such as periods (.), no additional annotation or introduction of new special tokens was necessary to handle sentence segmentation. During the model training phase, the model learned the distribution and usage patterns of periods in natural language through the language modeling task. Therefore, for sentence segmentation, it is sufficient to use the model to predict the probability that the next token in a given text is a period, thereby completing the sentence segmentation.

Experimental results demonstrated that this method can efficiently and accurately identify sentence boundaries, with generated sentence boundaries conforming to semantic logic. The approach also exhibited strong adaptability across various language scenarios. This unified strategy not only simplified model design but also significantly improved the efficiency of sentence segmentation tasks.

## 2.6 Summary model

**Data.** We chose the Amazon English Review Dataset as the source for training and testing data. This dataset includes user reviews and their corresponding review titles, where the review content is typically long text, and the titles provide a concise summary of the content. To construct a summarization task dataset, we used the review content as input text and the titles as target summaries. After preprocessing, we filtered out samples that were too short or contained excessive noise, ensuring high-quality training data and consistency in the task.

**Model Training.** For model selection, we again adopted the GPT2LMHeadModel, which is based on GPT-2 and specifically designed for causal language modeling tasks. Since GPT-2 does not directly support Seq2Seq problems, we indirectly achieved this by formatting the training samples. Specifically, each training sample was organized into the format "[Original Text] TL;DR: [Summary]", where "TL;DR:" serves as a prompt to guide the model in understanding the relationship between the input text and the target summary. During training, the model learned to generate concise summaries from long texts by maximizing the log-likelihood estimation of the training data, while preserving the core information of the input text.

**Model Usage.** In the deployment phase, to enhance the accuracy of the generated results, we introduced a dynamic masking mechanism. User inputs, such as DNA sequences, were formatted as "[Original Text] TL;DR:". However, during summary generation, we constrained the model's output. Specifically, by creating masks, we set the scores of tokens unrelated to DNA or relevant fields to -inf, ensuring that the model only outputs vocabulary pertinent to the DNA theme. This method filters out irrelevant candidate tokens, making the generated summaries more aligned with expected expressions.

## 2.7 Genome Segmentation - Chapter Division - Sentence Splitting

We utilized the human genome data GRCh38.p14, focusing primarily on processing Chromosome 1. Our approach involved dividing and structuring the chromosome into multiple hierarchical levels, as detailed below:

**Chromosome Part:**

We first divided Chromosome 1 into 25 segments, each containing approximately 10 MB of DNA sequence. This initial division is referred to as the "Chromosome Volume" level.

**DNA Paragraphs:**

For each segment, we applied a segmentation model to further divide the sequences into basic paragraphs. For example, the first segment resulted in a total of 15,238 paragraphs. This level of structure is termed "DNA Paragraphs."

**DNA Sections:**

Using a pre-trained model, we generated vector representations for each paragraph from Step 2. We then performed dynamic clustering on these vectors to form a second-level directory. The first segment resulted in a total of 503 sections at this level. This intermediate structure is referred to as "DNA Sections."

**DNA Chapters:**

Utilizing the centroid vectors from the second-level paragraphs, we executed another round of dynamic clustering to form third-level paragraphs. The first segment ultimately produced 38 chapters at this level. This higher-level structure is termed "DNA Chapters."

This process can be extended to create even higher levels of paragraph structures. However, since the 38 third-level paragraphs align well with the typical structure of a book, we limited our hierarchy to three levels.

As described above, the hierarchical structure of the genome catalog we constructed is as follows:

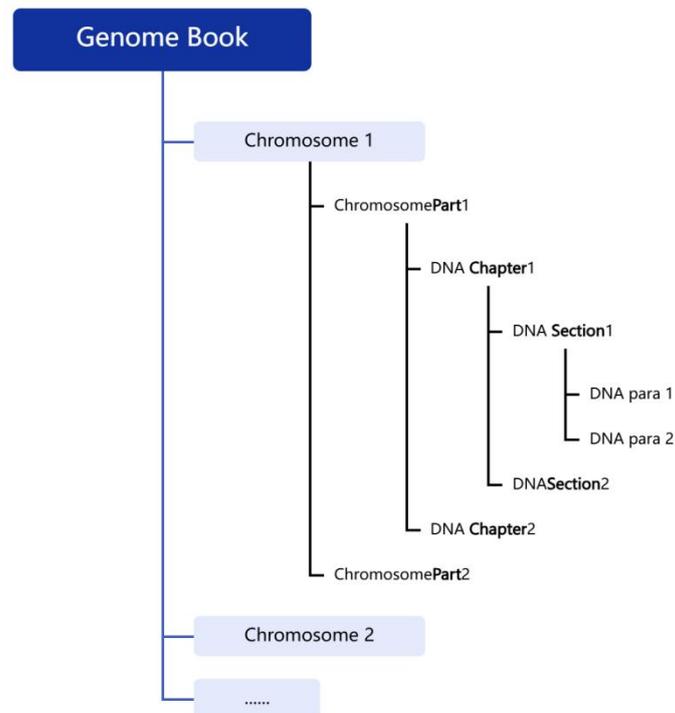

Fig.4. Structure of the Genome Catalog. Each genome is organized by chromosome, forming the highest-level directory. We divided the chromosome sequences into equal-sized segments, referred to as "Parts." For each part, we further subdivided into "Chapters" and "Sections," where each section represents specific DNA sequence paragraphs.

After completing the segmentation and chapter division, we used a sentence segmentation model to process each paragraph into sentences.

For generating titles for chapters and sections, we focused on creating titles for "DNA Sections" and "DNA Volumes":

**Title Generation for "DNA Sections":**

Extract summaries from all DNA paragraphs within the section.

Concatenate these summaries in order to form a long sequence.

Perform summary extraction on this concatenated sequence to obtain the title for the "DNA Section."

**Title Generation for "DNA Volumes":**

Similar to the above method, concatenate the titles of all "DNA Sections" within the volume to form a long sequence.

Perform summary extraction on this concatenated sequence to obtain the title for the "DNA Volume."

Most summarization algorithms require processing only a portion of the core text. In this case, our pre-trained model has a maximum input size of 256 tokens, meaning that the first 256 tokens of the sequence are used for summary extraction.

### 2.8 Translation of DNA to English

We used the pre-trained model gpt2-gene-eng-ft to generate vectors for all English words and DNA-related terms, utilizing the last hidden layer of the GPT-2 network. For each DNA term, we queried the English word with the highest vector similarity to serve as its corresponding English translation. Although fine-tuning brought the distances between DNA terms and English words closer, there remains a noticeable overall distance. Consequently, approximately 19,000 English words corresponded to about 600 unique English vocabulary items. It is important to note that this word translation relationship is based solely on structural similarity in the embedding space and does not imply semantic similarity.

After constructing the translation dictionary from DNA to English, we tokenized and translated each DNA paragraph sequence. This process allowed us to obtain an English version of the complete human genome book.

Generating partial content of the book, as shown in the screenshot below:

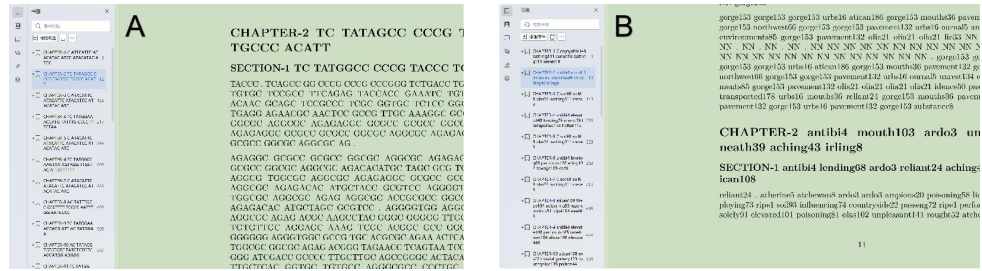

Fig. 5. Screenshot of the Genomic Book Content, where Figure A on the left shows the DNA sequence content, and Figure B on the right shows the translated English content.

# 3 Discussion

This paper employs a large language model to construct segmentation, sentence splitting, and summarization models for DNA sequences. We then applied these models to process the human genome, resulting in a fully structured genomic book that is divided into chapters, segments, sentences, and tokens. Based on this research, several novel DNA analysis applications may emerge:

**3.1 Fast DNA Search**

By segmenting DNA sequences into chapters, paragraphs, and sentences, a multi-level indexing structure can be constructed, similar to a hierarchical file system:

- **Chapter-Level Search**: Quickly locate target regions, reducing the search scope.
- **Paragraph-Level Search**: Achieve precise matching of specific sequence paragraphs.
- **Sentence-Level Search**: Identify minor variations within smaller segments.

This layered structure significantly enhances search efficiency, particularly when handling large-scale genomic databases.

**3.2 Genomic Unique Identifier**

Utilizing the Summarization Capabilities of Large Language Models to Generate Unique Feature Identifiers for Genomes

**Fault-Tolerant Identifiers**: The model-generated identifiers are robust to small variations, such as point mutations, ensuring that they can still recognize similar or identical genomes despite minor changes.

**Comparative Analysis**: The identifiers enable rapid comparison of similarities and differences between different genomes, supporting evolutionary analysis and disease research.

By leveraging the summarization capabilities of large language models, we can generate unique feature identifiers for genomes that offer both fault tolerance and efficient comparative analysis. This approach enhances the accuracy and utility of genomic studies.

### 3.3 Genome Data Compression and Storage Optimization

Based on Summarized or Translated Genomic Representations:

**Efficient Storage**: Reduce redundant data storage by compressing genomic information using chapter summaries.

**Interpretable Data Format**: Generate a more human-readable format for genomic data, facilitating quicker comprehension by scientists.

By utilizing summarized or translated genomic representations, we can achieve both efficient storage solutions and enhance the interpretability of genomic data, making it easier for researchers to understand and work with.

## 4 Conclusion

This paper leverages the multilingual transfer ability of large language models to construct a comprehensive representation of the human genome, organized into DNA "words," "sentences," and "paragraphs." This endeavor achieves a significant goal in genomics research and provides new perspectives for developing advanced DNA search technologies, storage methods, and other applications. If the analytical, summarization, and reasoning capabilities of current large language models can be effectively transferred to DNA analysis, it could unlock substantial technological advancements.

Since research on the interpretability of large language models is still in development, our constructed genomic book currently does not provide definitive biological interpretations. This study is primarily illustrative, showcasing the potential of applying natural language ability to DNA analysis. We aim to stimulate further thought and innovation in applying large language models to biological research, serving as a stepping stone for future advancements.